\documentclass[aip,apl,floatfix
,preprint
]{revtex4-1}
\usepackage[]{graphicx}
\usepackage{amsmath}
\usepackage{amssymb}
\usepackage{amstext}
\usepackage{gensymb}
\usepackage[version=4]{mhchem}
\usepackage[super]{nth}

\newcommand{\qz}{$\vec{q}_\text{z}$}
\newcommand{\qy}{$\vec{q}_\text{y}$}

\newcommand{\IA}{\AA$^{-1}$}

\renewcommand{\vec}[1]{\mathbf{#1}}
\let\oldhat\hat
\renewcommand{\hat}[1]{\oldhat{\mathbf{#1}}}

\usepackage{xcite}

\usepackage{xr}
\makeatletter
\newcommand*{\addFileDependency}[1]{% argument=file name and extension
  \typeout{(#1)}% latexmk will find this if $recorder=0 (however, in that case, it will ignore #1 if it is a .aux or .pdf file etc and it exists! if it doesn't exist, it will appear in the list of dependents regardless)
  \@addtofilelist{#1}% if you want it to appear in \listfiles, not really necessary and latexmk doesn't use this
  \IfFileExists{#1}{}{\typeout{No file #1.}}% latexmk will find this message if #1 doesn't exist (yet)
}
\makeatother

\newcommand*{\myexternaldocument}[1]{%
    \externaldocument{#1}%
    \addFileDependency{#1.tex}%
    \addFileDependency{#1.aux}%
}
\myexternaldocument{SI}
\begin{document}

%\preprint{}

\title{X-ray reflectivity with a twist: quantitative time-resolved X-ray reflectivity using monochromatic synchrotron radiation} %Title of paper
%\title{An improved geometry for quantitative time-resolved x-ray reflectivity using monochromatic synchrotron radiation} %Title of paper

% ARW notes on the title: "An improved geometry" is accurate but not as inspirational as one could wish.
%\title{X-ray reflectivity with a twist} %Title of paper

\author{Howie Joress}
\email[]{hj335@cornell.edu}
\affiliation{Cornell High Energy Synchrotron Source, Cornell University, Ithaca, NY 14853}
%\affiliation{Department of Materials Science and Engineering, Cornell University, Ithaca, NY 14853}
\author{Shane Arlington}
\affiliation{Department of Materials Science and Engineering, Johns Hopkins University, Baltimore, MD 21218}
\author{Timothy P. Weihs}
\affiliation{Department of Materials Science and Engineering, Johns Hopkins University, Baltimore, MD 21218}
\author{Joel D. Brock}
\affiliation{Cornell High Energy Synchrotron Source, Cornell University, Ithaca, NY 14853}
\affiliation{School of Applied and Engineering Physics, Cornell University, Ithaca, NY 14853}

\author{Arthur R. Woll}
\affiliation{Cornell High Energy Synchrotron Source, Cornell University, Ithaca, NY 14853}

\date{\today}

\begin{abstract}
We have developed an improved method of time-resolved x-ray reflectivity (XRR) using monochromatic synchrotron radiation.   Our method utilizes a polycapillary x-ray optic to create a range of incident angles and an area detector to collect the specular reflections.  By rotating the sample normal out of the plane of the incident fan, we can separate the surface diffuse scatter from the reflectivity signal, greatly improving the quality of the XRR spectra compared to previous implementations.  We demonstrate the time-resolved capabilities of this system, with temporal resolution as low as 10 ms, by measuring XRR during the annealing of Al/Ni nano-scale multilayers and use this information to extract the activation energy for interdiffusion in this system.

\end{abstract}


\maketitle %\maketitle must follow title, authors, abstract and \pacs

%\section{Introduction}

X-ray reflectivity (XRR) is a well-known method for measuring the structure of interfaces and thin-films with both high accuracy and atomic-scale precision.\cite{als2011elements}  Like many x-ray scattering methods, XRR is particularly attractive for in-situ studies because it is  a remote probe and can be performed in the presence of environments including atmospheric pressure and liquids.\cite{kowarik_rev} On the other hand, applications of XRR to time-resolved processes is limited. Because XRR is typically performed by scanning both the sample and detector,\cite{Parratt} a single scan usually requires several seconds or minutes,\cite{mocuta2018fast,lippmann2016new} depending on the details of the setup and scan parameters. 

Although many variations of XRR, referred to broadly as quick XRR (qXRR),\cite{sakurai} have been developed to overcome this limitation, they generally place severe constraints on the source and/or sample which make them incompatible with certain applications. For example, one approach to qXRR, known as convergent beam x-ray reflectivity (CBXR),\cite{naudon1,miyazaki,stoev} permits the use of monochromatic radiation, but generally requires a highly diverging \textit{incident} beam, and thus cannot exploit the high intensity, but also highly collimated, beams provided by synchrotron sources.  Recently, we demonstrated a novel approach to CBXR that makes use of a polycapillary (PC) optic to overcome this restriction and permit the use of monochromatic synchrotron radiation.\cite{Joress:hf5357} Although this method greatly broadens the potential scope of qXRR, our implementation, similar to other implementations of CBXR, is limited due to the overlap of the sharp XRR signal with the diffuse scatter from the surface. 

Recently, making use of a lab-based source, \citet{Voegeli:rg5123} demonstrated that the resolution obtained with CBXR can be improved by rotating the sample such that the surface normal is out of the plane of the incident beam. Here we demonstrate the combination of this low-background geometry with our PC-based method\cite{Joress:hf5357} at the Cornell High Energy Synchrotron Source (CHESS).  We demonstrate that we can record, on a single detector image, a portion of the reflectivity curve in good agreement with traditional Parratt XRR and with time resolution of as low as 10 ms.  We also demonstrate this technique by showing qXRR during the annealing of an Al/Ni multilayer (ML) with detector framing  at 9 Hz.

\begin{figure*}
\includegraphics{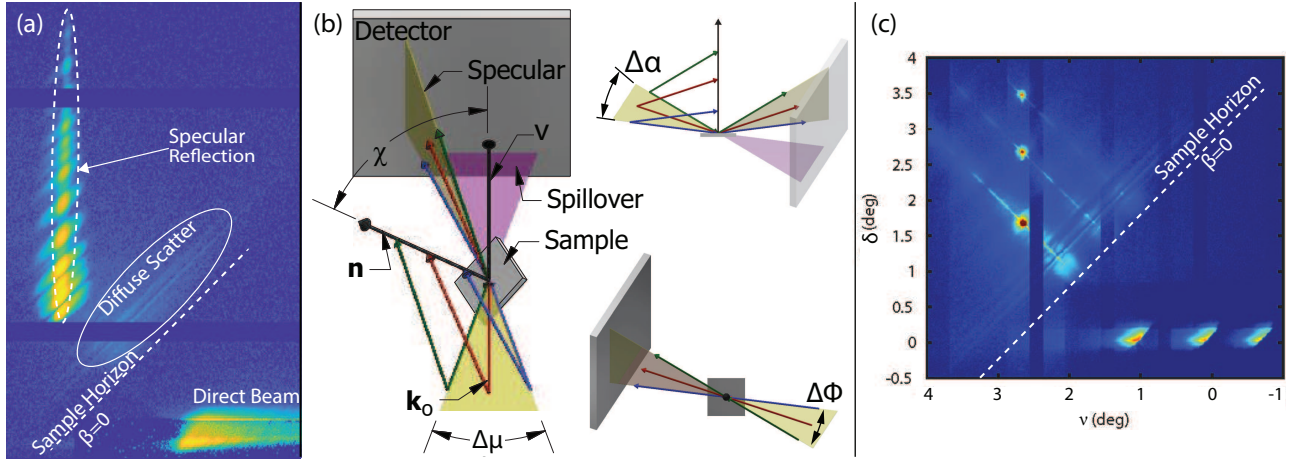}%
\caption{\label{fig1comp}(a) A portion of a detector image showing qXRR (b) Schematic representations of the scattering geometry with three incident vectors drawn explicitly. 
Clockwise from the left the schematics are viewed from above the incident beam, parallel to the substrate surface, and along $\hat{n}$. (c) A composite of three separate detector images, each obtained with a single, collimated incident beam.  They are combined to simulate data obtained using an incident beam fan consisting of three discrete incident rays rather than a continuous fan. The spillover from each incident beam is visible at the lower right. In (a) and (c) positive $d_\text{h}$ is positive to the left and  the image intensities  are on a log scale to improve visibility of the diffuse as well as specular intensity.    %The specular reflections all fall in a vertical line and the diffuse scatter is primarily confined to the scattering plane.  
The dark horizontal bands in (a) and vertical bands in (c) are due to "dead space" between detector modules.}%

\end{figure*}
The qXRR measurements were primarily performed at CHESS, primarily the G3 station.\footnote{The G1 hutch was also used for some measurements.  Other than small differences in energy and the sample-to-detector distance, these set-ups were nominally identical.}  The experimental geometry is shown schematically in Fig \ref{fig1comp}(b).  A PC x-ray optic\cite{polycap1} is used to create a converging fan of radiation in the horizontal plane, with an angular range of $\Delta\mu\approx 2.5\degree$.   The sample is placed at the focal point of this beam.  For these experiments everything upstream of the sample is identical to that described in Ref. \citenum{Joress:hf5357} with the exception of a vertical slit placed just upstream of the sample, added to reduce the vertical divergence of the incident fan out of the PC. The sample normal, $\hat{n}$, is tilted out of the plane of the incident beam fan by angle $\psi=90\degree-\chi$, so that the incident rays %, $\vec{k}_\text{in}$, 
strike the sample with a range of incident-beam angles, $\Delta\alpha=\Delta\mu \sin(\chi )$, relative to the sample surface and a range of azimuthal angles, $\Delta \phi$, about $\hat{n}$ (see diagrams at the upper and lower right of Fig \ref{fig1comp}(b)). 
Experimentally, this is achieved by using a goniometer (shown schematically in SI Fig. \ref{goni_drawing}) that comprises two rotational stages: a semicircle (Huber 5202.40) with a horizontal rotational axis, acting as the $\chi$ degree of freedom, on top of a rotation stage (Huber 410) with a vertical axis, the $\theta$ degree of freedom used to set $\alpha_0=\arcsin{\left[\sin{\theta}\sin\chi\right]}$.  For this work we use $\chi=45\degree$.  The two stages are aligned with each-other such that the center of rotation of the upper stage was along the rotational axis for the lower stage.  These two stages are then aligned using a set of linear stages, vertical and transverse to the beam, such that the focal point of the focused beam is coincident with the center of rotation of the goniometer. %intersection of the rotational axes of the two rotary stages.  
The sample is mounted such that the surface is parallel to the $\chi$ rotation axis.  A linear translation stage is used to move the sample surface to the center of rotation.
 Following the usual conventions, we define $\alpha$ and $\beta$ as the angles between the sample surface and each incident beam vector, $\vec{k}_\text{in}$, and exit beam vector, $\vec{k}_\text{out}$, respectively, and their difference as $\Omega=\beta-\alpha$. The central ray of the incident fan, $\vec{k}_0$, is at angle $\alpha_0$ from the sample surface.  An area detector is placed downstream to collect the diffracted intensity.  For these measurements, we used a low-noise, photon counting area detector (Dectris Pilatus 300k), placed around 1.8 m downstream of the sample ($D_{\text{SD}}$).  It is positioned by means of vertical and horizontal linear translation stages.  Fig. \ref{fig1comp}(a) shows a portion of a detector image.  $\nu$ and $\delta$ refer,  respectively, to the horizontal and vertical scattering angles measured from $\vec{k}_0$. Given that $\chi=$45\degree{}, the specular condition is met for all rays in the incident fan when $\nu_\text{spec}=2\alpha_0/\sqrt[]{2}$. The specular reflections, which form the bright vertical feature on the detector, appear 90\degree{}  from the plane of the incident fan (spillover from which is seen on the lower right of the detector image). 

As discussed by \citet{Voegeli:rg5123}, this geometry separates the specularly-reflected beam from the brightest portion of the diffuse scattering from the surface, which is confined to planes, which we will refer to as scattering planes, defined by the incident beam vector and $\hat{n}$, as shown in Fig. \ref{fig1comp}(b). This effect, the origins which are described in SI Section \ref{res_der}, is illustrated in Fig \ref{fig1comp}(c): the image is a sum of three separate measurements obtained with the same sample and detector geometry as in Fig \ref{fig1comp}(a), but with a collimated incident beam at three different $\mu$ angles.  %, rather than a continuous converging beam. 
This image thus simulates a single, qXRR snapshot obtained with three, discrete converging beams rather than a continuous fan. As in Fig \ref{fig1comp}(a), the specular reflections are along a vertical line.  Emanating from each of the specular features at a 45\degree{} angle is a line of diffuse scatter which falls along the scattering plane belonging to each $\vec{k}_\text{in}$. 

In order to extract XRR curves from images, such as Fig.\ref{fig1comp}(a),  we make use of the above observation.  We ignore contributions to scattering  that do not fall along the scattering plane corresponding to their respective incident ray. Specifically, we assume that each 45\degree{} line parallel to those in Fig \ref{fig1comp}(c) corresponds to a unique value of $\alpha$ and a range of values of $\Omega$. Then, every point on the detector can addressed with a unique \qz{} and \qy{} based on its coordinates, $d_\text{h}$, horizontal, and $d_\text{v}$,vertical, relative to the location where $\vec{k}_0$ intercepts the detector plane (diffraction is in the positive $d_\text{h}$ and $d_\text{v}$ directions):
\begin{align}
\vec{q}_\text{z}=k\left[\cos(\alpha+\Omega)-\cos(\alpha)\right],
\\
\vec{q}_\text{y}=k\left[\sin(\alpha+\Omega)+\sin(\alpha)\right],
\end{align}
where
\begin{align}
2\alpha&=\arctan\left(\sqrt[]{2}\left[d_\text{v}-(d_\text{h}-d_{\text{h,spec}})\right]/D_{\text{SD}}\right), \textnormal{and} \label{eq:2a}
\\
\Omega&=\arctan\left(\sqrt[]{2}d_{v}/D_\text{SD}\right)-2\alpha.\label{eq:omega}
\end{align}
$d_{\text{h,spec}}$ is the horizontal specular position on the detector ($d_{\text{h,spec}}=D_\text{SD}\tan(\nu_\text{spec})$, constant for each image), and $D_\text{SD}$ is the distance from the sample to the detector plane.

The process of qXRR extraction begins by normalizing the detector images to the average incident intensity (for time resolved data).  For some measurements a Si wafer was used to attenuate the scatter on a portion of the detector, extending its effective dynamic range; if applicable, we then correct for this partial attenuation.  Images are additionally corrected for the variation in intensity as a function of $\alpha$.  This variation was measured by imaging the direct beam from the PC. The correction was performed under the assumption that 45\degree{} lines on the detector have a constant $\alpha$ angle as described above.  %By collecting and measuring the intensity along the far-field pattern of the direct beam downstream of the PC, we determine the relative incident flux as a function of $\alpha$.  %The image can then be normalized to the incident flux using the assumption that lines along the detector have a constant $\alpha$ angle as described above.   
%Images are additionally corrected for this intensity variation, assuming that each 45\degree{} line on the detector corresponds to a single $\alpha$-angle  as described above.  
Using the algorithm in Ref. \onlinecite{barna1999calibration}, we then re-map the data from linear detector coordinates to $2\alpha$ and $\Omega$ space according to Eq. \ref{eq:2a} and \ref{eq:omega} (see SI Fig. \ref{transform}). In this coordinate system the diffuse streaks are  perpendicular to the specular direction. This is a more convenient coordinate system to work in, as opposed to remapping to \qz{} and \qy{} space, as the detector remaps to a relatively equally spaced grid. For data where there is a lot of background or diffuse scatter, background subtraction is performed by applying a linear fit to the background, line-by-line, along the specular direction.  Finally, the reflectivity curve is generated by summing across the width of the specular feature, and $2\alpha$ is converted to \qz{}.

\begin{figure}
\includegraphics[]{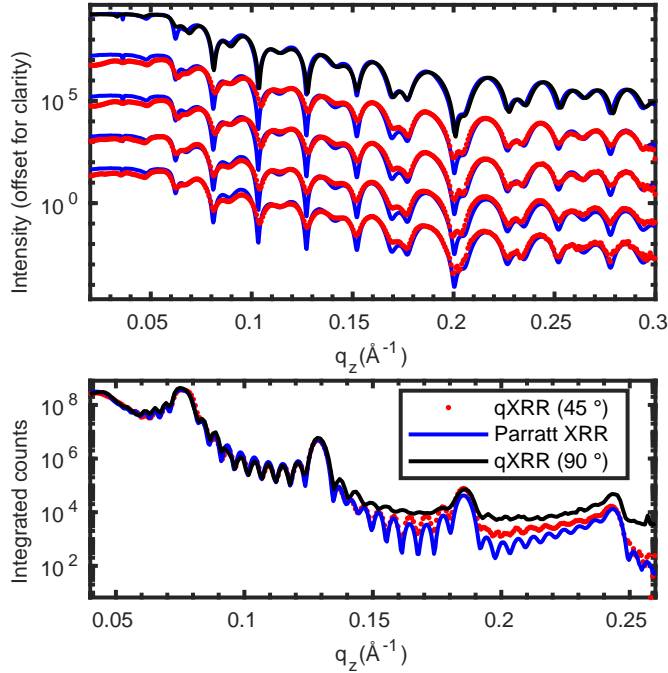}
\caption{\label{qxrrcomp}Comparison of qXRR with Parratt reflectivity from two samples.  (a) is from a trilayer film.  Continuous blue lines shows Parratt reflectivity.  The top plot (black line) shows a model fit to the Parratt data and the four lower plots, from  top to bottom show qXRR taken with 10 s, 1 s, 100 ms, and 10 ms respectively. (b) shows XRR from a 10 bilayer Al/Ni film with 105 \AA{} periodicity shown taken using Parratt XRR along with qXRR taken at $\chi=45\degree \textnormal{ and } 90\degree$.}
\end{figure}

To demonstrate the accuracy of our measurement method, we show XRR from two different samples, both by traditional Parratt XRR as well as the qXRR method described here.  The Parratt reflectivity measurements were taken at the CHESS G2 station with the diffraction measurement done in the vertical direction.  X-rays are fed to the station from a Be side bounce monochromator with 0.1\% $\Delta E/E$ bandpass, and the beam is slit down to $<0.1$ mm in the vertical direction.  Reflectivity was recorded using a 1D strip detector (Dectris Mythen 1K) with the detector strip in the vertical direction to enable background subtraction.    %The Parratt XRR shown here was measured at the CHESS G2 hutch.  
Fig. \ref{qxrrcomp}(a) shows reflectivity from a trilayer film, grown by the optics group of the Advanced Photon Source, containing nominally 200 \AA{} of Mo, 200 \AA{} of \ce{B4C}, and a capping layer of 67 \AA{}  of Mo.    We show qXRR collected over 4 decades of integration time, from 10 s to 10 ms.  The Parratt XRR was fit to a model using GenX.\cite{Bjorck:aj5091} It suggests that the layers are 195, 250, and 33 \AA{} respectively.  For the 10 s and 1 s curves there is good agreement between the two XRR curves other than in the lowest intensity portions of the curves in the troughs between fringes.  When collecting the 100 ms and 10 ms data, a piece of Si wafer was placed in front of part of the detector in order to extend its effective dynamic range by attenuating the brightest part of the diffracted intensity.  These curves have some additional background, though significant noise is only found at the lowest intensity parts of the 10 ms curve.

In Fig. \ref{qxrrcomp}(b) we show XRR from a 10 bilayer thick Al/Ni ML film on SiC.
The MLs were deposited in a custom deposition system.\cite{joress2012self}  The films were grown from an Al 1100 alloy target and a Ni target with 7 wt\% V.  The MLs were fabricated with an aim to achieve a 1:1 atomic ratio of Al to Ni-V which results in a 3:2 thickness ratio for the alternating layers.  The films were grown on 6H-SiC wafers with a on-axis c-plane surface.  Again, the qXRR with $\chi=45\degree$ agrees well with the Parratt reflectivity.  For this film, with a bilayer thickness of 105 \AA{}, the qXRR has resolution sufficient to capture the Kiessig fringes throughout the entire range of the measurement.  There are some differences in absolute intensity of some of the Kiessig fringes past the \nth{2} ML peak though the intensity of the ML peaks are well captured.  Between the \nth{3} and the \nth{4} ML peaks the Kiessig fringes are still visible but are damped due to higher background.  For comparison, qXRR at  $\chi=90\degree$ is also shown (extraction of qXRR in this geometry is described in detail in Ref. \onlinecite{Joress:hf5357}).  Again the qXRR matches the Parratt XRR through the \nth{2} ML peak.  At higher \qz{} there is a large amount of additional signal intensity due to diffuse scattering.  Between the \nth{3} and \nth{4} ML peaks the Kiessig fringes are largely inseparable from the noise. 
Because of the geometry, our new method does not suffer from this same degradation of the XRR signal.
An additional example of this effect is shown in SI Fig. \ref{10per}.  With changes in the experimental implementation, such as using an optic with a shorter WD, improvements including reducing requirement on incident beam width, increasing $\Delta\alpha$, and improved \qz-resolution can be realized.

\begin{figure}
\includegraphics[]{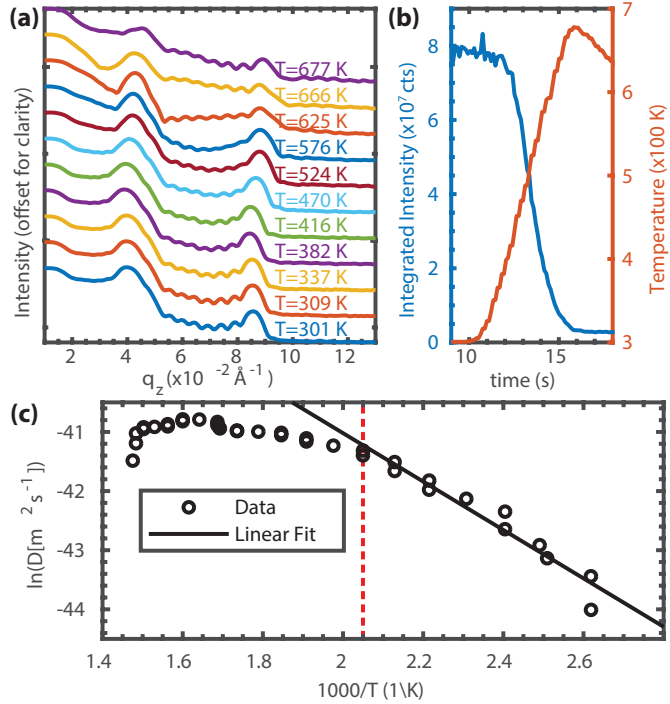}
\caption{\label{trcomp}(a) Time resolved qXRR during heating of an Al/Ni ML at up to a rate of 90.9 K/s.  The \nth{1} ML peak at ~0.4 \IA{} decreases due to intermixing.  Frames were taken at ~9 Hz and every \nth{5} frame is shown.  Temperature at each point is shown on the right.  (b) Extracted intensity of the \nth{1} ML peak and temperature as a function of time. (c) An Arrhenius plot derived as described below.  A linear fit, used to extract $E_0$, based on data taken at $1000/T>2.05\;K^{-1}$ is also shown.}
\end{figure}
To demonstrate the time resolved ability of this technique, we measured qXRR from Al/Ni MLs on SiC while heating from room temperature to above 575 K at rates of up to 90.9 K/s.  This was achieved by incorporating a custom designed low thermal mass heating stage into the goniometer  (see SI Fig. \ref{heater}).  The heater itself is an aluminum nitride-based heating plate with output of up to 580 W at temperatures up to 400 \degree{}C. The heater was mounted in a Macor ceramic block.  The samples are held onto the heater by 316 stainless steel clips, which are insulated from the sample by pressed mica sheet.  The clips are separated from each other by about 7 mm.  A foil thermocouple is placed between the sample surface and the mica insulation on one side to measure the temperature of the film.  Al and Ni have a negative enthalpy of mixing and therefore, when the film is sufficiently hot, the films intermix relatively rapidly.  This intermixing can be readily measured by XRR.  Fig. \ref{trcomp}(a)  shows the evolution of the qXRR as a function of time during the heating and Fig. \ref{trcomp}(b) shows the intensity of the \nth{1} ML reflection, $I$,  and temperature, $T$, as a function of time, $t$.  Frames were recorded with 100 ms integration time with the detector framing at ~9 Hz.  As the sample is heated there is a loss of intensity of the ML peak due to weakening of the periodicity of the film and therefore the Fourier amplitude.  It should also be noted that there is no change in the frequency of the Kiessig oscillations throughout the measurement since there is little change in the thickness of the film.  There is some change, however, in the shape of the qXRR curves -- for instance, the change in slope between the \nth{2} and \nth{3} ML peaks at 625 K. We attribute this to the formation of a surface oxide,  resulting in a low frequency Kiessig oscillation.

By looking at  $I(t)$ over time, we can extract the diffusivity.  \citet{dumond1940x} derived an equation relating the relative intensity of a ML reflection with the interdiffusion constant, $\tilde{D}$, during constant heating:
\begin{equation}
\tilde{D}=\frac{-\Lambda^2}{8\pi^2}\ln\left[\frac{I(t)}{I_0}\right]\frac{1}{\Delta t}.
\end{equation}
Here $\Lambda$ is the bilayer spacing and $I_0$ is the initial intensity.  %However, because the films are very thin, the intermixing   
We can apply this equation to our measurements during heating by using its derivative form,\cite{wang1999interdiffusion,greer_spaepen_1985} 
\begin{equation}
\tilde{D}=\frac{-\Lambda^2}{8\pi^2} \, \frac{d}{d t}\ln\left[\frac{I(t)}{I_0}\right],
\end{equation}
treating each point during heating as an infinitesimally short isothermal diffusion period.  To apply a derivative to relatively discrete data, we applied a smoothing spline\cite{matlab} (P=0.99) to the extracted $I(T)$.  The resulting interdiffusion constants are plotted in Fig. \ref{trcomp}(c) as a function of temperature. As seen in the plot, at low temperatures the diffusivity has an Arrhenius relationship with temperature.  Fitting this low-temperature region gives an activation energy, $E_0$, of 34.0 kJ/mol.    We attribute the deviation from Arrhenius behavior at higher temperatures and longer times to a decrease in the rate of intermixing as the system approaches equilibrium and the concentration gradient between layers decreases.  This effect should be mitigated, increasing the accuracy of the measurment, by increasing the bilayer thickness or increasing the temperature ramp-rate, maintaining the concentration gradient through higher temperatures.  We then repeated the previous measurement 4 more times with a range of heating rates down to 34.5 K/s  and  found the results to be relatively consistent regardless of ramp rate (the data can be found in SI Fig. \ref{all_arrhenius} and Table \ref{E0res}).  The average $E_0$ was $36.3\pm 2.50$ kJ/mol. Ref. \onlinecite{fritz2013thresholds} has a table of experimentally measured $E_0$ values obtained using a variety of methods by a variety of research groups; they vary from 17.4 to 523 kJ/mol.   

In summary, we have demonstrated an improved geometry for qXRR, with data quality suitable for quantitative analysis at acquisition times reaching 10 ms.
We then applied this approach to the study of intermixing in metal-metal nano-scale MLs.  With a time resolution of 100 ms, we recorded XRR curves during annealing of the MLs and were able to generate Arrhenius curves for each reaction.

%\begin{acknowledgments}
This research was funded by and conducted at the Cornell High Energy Synchrotron Source (CHESS) which is supported by the National Science Foundation under award DMR-1332208.  The authors thank the CHESS technical staff for their assistance, particularly J. Houghton, C. Bagnell, J. Hopkins and E. Edwards for assisting in constructing components of the experimental set-up.  We thank R. Conley and A. Macrander (Advance Photon Source) for growing the tri-layer film.  We also thank T. Hufnagel (Johns Hopkins) for his useful discussion and loaning us proof-of-concept films.
%\end{acknowledgments}

\bibliography{refs.bib}

\end{document}